\documentclass[english,prl,floats,twocolumn,superscriptaddress,showpacs]{revtex4-1}
\usepackage[T1]{fontenc}
\usepackage[latin9]{inputenc}
\usepackage{bm}
\usepackage{amsmath}
\usepackage{amssymb}
\usepackage{graphicx}
\usepackage{bm}
\usepackage{ dsfont }

\def\to{\rightarrow}

\def\p{\partial}

\def\s{\sigma}

\newcommand{\beq}{\begin{equation}} 
\newcommand{\eeq}{\end{equation}} 
\newcommand{\beqa}{\begin{eqnarray}} 
\newcommand{\eeqa}{\end{eqnarray}}


 
\usepackage{babel}
\begin{document}
\bibliographystyle{apsrev4-1}

\title{Rotational strain in Weyl semimetals. A continuum approach.}
\author{Vicente Arjona}  
\affiliation{ Instituto de Ciencia de Materiales de Madrid, and CSIC, Cantoblanco, 28049 Madrid, Spain}
\author{Mar\'ia A. H. Vozmediano}
\affiliation{ Instituto de Ciencia de Materiales de Madrid, and CSIC, Cantoblanco, 28049 Madrid, Spain}

\begin{abstract}
We use a symmetry approach to derive the coupling of lattice deformations to electronic excitations in three dimensional Dirac and Weyl semimetals in the continuum low energy model. We focus on the effects of rotational strain and show that it can drive transitions from Dirac to Weyl semimetals, gives rise to new elastic gauge fields, tilts the cones, and generates new pseudo--Zeeman couplings. It also can generate a deformation potential in volume--preserving deformations. The associated pseudo--electric field contributes to the chiral anomaly. 
\end{abstract}
\maketitle

Dirac and Weyl semimetals (WSMs) \cite{Liuetal14,Letal14b,Xu15,Lvetal15} are three dimensional crystalline materials whose band structure shows nontrivial band-crossings close to the Fermi level. Their low energy excitations  around these points are described by a Weyl Hamiltonian \cite{Weyl29} $H_{\pm}=\pm v_F{\vec \sigma}\cdot{\vec p}$ and realize the high energy predictions of chiral physics:  giant negative magnetoresistance as a consequence of the chiral anomaly \cite{KK13,XKetal15,Lietal15,ZXetal16}, chiral magnetic effect \cite{LKetal16} and even signatures of mixed gravitational anomaly \cite{GNetal17}.   

The locations of the Weyl points in momentum space determine the character and stability of the material. Dirac semimetals, the first realized experimentally, have the chiral pairs located at  the same point in the BZ. Weyl semimetals (WSMs) have the Weyl nodes separated either in energy (inversion (${\cal I}$) breaking) or in momentum (time-reversal (${\cal T}$) broken). They are classified into type I and type II depending on the tilt of the Weyl nodes \cite{SGetal15}. Strain has a strong influence on the band structure of topological semimetals and can drive transitions between them \cite{ZJetal13,SRetal16,RJetal16}. In this work we use a symmetry approach to derive effective low energy electron--phonon  Hamiltonian  interactions in Dirac and WSMs. In particular, we will see that lattice distortions can induce transitions between type I and type--II WSMs and rotational strain is able to make transitions from Dirac to  Weyl semimetal.

The minimal low energy effective Hamiltonian describing a WSM with only two nodes of $\pm$ chiralities separated in energy-momentum space by a vector $(b^0,\vec{b})$ is
\beq
H_{\pm}=\pm[{\vec\sigma}({\vec k}\pm{\vec b})]\pm b_0\sigma_0,
\label{HWSM}
\eeq
where ${\vec\sigma}$ are the  Pauli matrices and $\sigma_0$ is the identity.
%
The vector ${\vec b}$ is an important intrinsic parameter of the WSM. 
The spacial components break ${\cal T}$ while the time component $b_0$ shifts the cones to different energies and breaks  ${\cal I}$ \cite{CFetal10}.

The deformation of a medium is described in elasticity theory \cite{LL71b} by the displacement ${\vec u}$ of a point at position ${\vec x}$ relative to an arbitrary origin ${\vec x_0}$. The deformation gradient tensor (or displacement gradient) ${\hat u}_{ij}$ is defined by the first order term in a Taylor expansion:
$u_i({\vec x})= u_i({\vec x_0})+\hat{u}_{ij}dx_j$, ${i,j}=\{1,2,3\}$, $\hat{u}_{ij}=\partial_i u_j$. For infinitesimal displacements, $\vert\hat{u}_{ij}\vert<<1$, the symmetric part defines the strain tensor 
\beq
u_{ij}=1/2(\partial_i u_j+\partial_j u_i),
\label{u}
\eeq
where $u_i$ is the  phonon vector field. Within elasticity theory, each derivative of the deformation gradient tensor is suppressed by a factor of order $\mathcal O (a/\lambda)$, where $\lambda$ is the wavelength of the deformation and $a$ is the lattice constant.

Effective low energy interactions between lattice deformations and electronic excitations are organized in derivatives of the electron field and of the deformation gradient tensor. Any term respecting the symmetries of the system is allowed. In the WSM  case the Weyl nodes are often located at points with no particular symmetry. We will restrict ourselves to this case and perform a construction based on the continuum model where the only symmetries are rotations in the plane perpendicular to the vector ${\vec b}$. We will construct effective Hamiltonians such that 
\beq
H=H_0+\sum_i g_{i}H_i+\sum_i {\tilde g}_i{\tilde H}_i,
\label{eph}
\eeq
where $H_0$ is given in eq. \eqref{HWSM}, and $H_i$ are given in tables \ref{T2} and \ref{tH}. We have separated the couplings associated to the antisymmetric components of the strain gradient tensor (Table \ref{T2}) which are the main contribution in this work, from these associated to the symmetric part (the standard strain tensor) (Table \ref{tH}). The last are similar to these extracted for graphene in ref. \cite{MJSV13} and will be  included at the end of the letter for completeness and to show the differences arising from the antisymmetric components.
Coupling constants remain arbitrary in this symmetry construction.  They have to be computed with a microscopic model like the  tight binding presented in \cite{SHR15,CFLV15}, calculated {\it ab initio}, or fixed by experiments.

In three dimensions, the antisymmetric part of the gradient deformation tensor $\omega_{ij}=1/2(\partial_i u_j-\partial_j u_i)$ is related to infinitesimal rotations represented by a vector ${\vec\Omega}$ by the expression
$\omega_{ij}= \epsilon_{ijk}\Omega_k$,
which can be inverted to give
$\Omega_k\equiv 1/2\epsilon_{ijk}\omega_{ij}$.
It is easy to see that the vector $\Omega_k$ is related to the deformation vector $u_j$ by
\beq
\Omega_k=\frac{1}{2}[\vec{\nabla}\times {\vec u}]_k.
\label{eq_rot}
\eeq
In fluid mechanics where ${\vec u}$ represents the fluid velocity, ${\vec \Omega}$ is the vorticity vector. This pseudo-vector is invariant under parity (inversion in 3D) and will couple with opposite sign to the two chiralities similarly to the vector ${\vec b}$. Contrary to the standard pseudo--gauge field ${\tilde H}_{2i}$ \cite{CFLV15} in table \ref{tH}, the vector $\Omega_i$ does not depend on the separation of the Weyl nodes in momentum space and its couplings will also affect Dirac semimetals. 
Rotational strain provides  two extra vectors to construct effective Hamiltonians in the symmetry approach.  The rotation vector itself ${\vec \Omega}$ will generate couplings affecting equally to Dirac and Weyl semimetals. ${\cal T}$--broken WSMs, will also form couplings through the vector: ${\vec b}\times{\vec \Omega}$.
 The lowest order terms in a derivative expansion are written in table \ref{T2}. 
\begin{widetext}
%
\begin{center}
\begin{table}[h]
\begin{tabular}{|c || c |c | c |c |}
\hline
$H_i$ & $(n_q, n_k)$ & Interaction term & Physical interpretation & $K_2$ \\
\hline\hline
$H_1$ & (0,0) & ${\vec \Omega}\cdot{\vec b}$ & Deformation potential & $+$  \\
  $H_{2}$ & (0,0) & ${\vec \sigma}\cdot{\vec\Omega}$,   ${\vec \sigma}\cdot({\vec b}\times{\vec\Omega})$ & Pseudogauge potentials  & $-$ \\
\hline
$H_{3}$ & (0,1) &  ${\vec\Omega}\cdot{\vec k}$,    $({\vec b}\times{\vec\Omega})\cdot{\vec k}$& Dirac cone tilts & $-$ \\
$H_{4}$ & (0,1) & $\omega_{ij} \s_i k_j $& It is zero & $+$ \\
 \hline
$H_5$ & (1,0) & $\;\;\;\;\epsilon_{ijk}\sigma_ik_jA_k\;\;\;\;$ & Pseudo-Zeeman term & $-$ 
\\
\hline
\end{tabular}
\caption{Lowest order effective low energy Hamiltonians for the electron-strain interactions around a 3D Weyl node associated to the antisymmetric part of the gradient deformation tensor.}
\label{T2}
\end{table}
\end{center}
\end{widetext}
The second column in the table indicates the number of derivatives acting respectively on the strain tensor  $(n_q)$ and on the spinor wave function $(n_k)$. The effective Hamiltonians are referred to one of the Weyl points. The column $K_2$ indicates the relative sign that the terms would have in an effective Hamiltonian built around the other chirality.


{\it Scalar potential}.
The term 
$H_{1}(u)={\vec \Omega}\cdot{\vec b}$, is a rotational contribution to the deformation potential. 
It  implies that a scalar potential can be generated from volume preserving deformations if local rotations are involved. We will later give a physical example of this situation. 

{\it Vector potentials}.
Rotational strain will generate two new axial gauge fields. 
The first coupling Hamiltonian 
$H_{2}=\sigma_i\Omega_i$ will affect the Weyl nodes separation.  
Being independent of ${\vec b}$, it implies that rotational strain can make a transition from Dirac  into WSMs similar to that discussed in \cite{GMS13}.
The second vector potential involves the Weyl nodes separation $b_i$ and is the analogue to  the one in ${\tilde H}_{2i}$ of Table~\ref{tH}: $A^{a}_i=\omega_{ij}b_j$. Using the definition of ${\vec \Omega}$, this vector can be written as
$A^{a}_i=\omega_{ij}b_j=\epsilon_{ijk}\Omega_k b_j=({\vec b}\times{\vec\Omega})_i$.

${\it Tilt}$.
A tilt of one cone is described in the low energy Hamiltonian as $H_T=W_ik_i$ where ${\vec W}$ is the tilt velocity. This term also breaks  ${\cal T}$ and it is another intrinsic parameter characterizing a given WSM. 
We have omitted so far the Fermi velocity. In the most general case, the Hamiltonian around one node will have the form
\beq
H=\sum_{i,j}v_{ij}k_i\sigma_j+(W_ik_i),
\label{Htilted}
\eeq
where $i,j=\{1,2,3\}$.
The dispersion relation is: 
\beq 
E(\mathbf{k})=\sum_i W_i k_i \pm \sqrt{\sum_j\left(\sum_i  v_{ij} k_i \right)^2}.
\eeq 
When  
\beq
(\lVert \mathbf{W} \rVert / \lVert \mathbf{v} \rVert ) >  1 ,
\label{1-2}
\eeq
the linear dispersion relation along the direction of the tilt will be overturned, the system develops a finite density of states and the Weyl semimetal becomes Type-II \cite{SGetal15}.

Both vectors $\Omega_i$ and $({\vec b}\times{\vec\Omega})_i$ induce  opposite tilts  in the two nodes. As before, being independent of the Weyl node separation, the coupling $\Omega_i k_i$ will tilt the cones in any Dirac material not protected by lattice symmetries. 

${\it H_{4}}$.
An interesting observation concerns the coupling $H_{4}=\omega_{ij}\sigma_ik_j$, involving the antisymmetric part of the deformation gradient tensor that would affect the Fermi velocity. As it was shown in ref. \cite{MJSV13}, the antisymmetric contribution to the Fermi velocity vanishes in the 2D case. We will se that it also vanishes here but for different reason. Consider the generic Hamiltonian 
$H=\epsilon_{ijk}\sigma_i k_ j V_k$,
where $\epsilon_{ijk}$ is the 3D Levi-Civita tensor and $V_k$ is an arbitrary vector field (it can be a constant). Due to the anticommutation relations $\sigma_i \sigma_j = i \epsilon_{ijk} \sigma_k + \delta_{ij}$, and using the symmetric convention for the derivatives acting on the electron fields, \hbox{$\psi^\dagger k_i\psi\!\to\! -i/2(\psi^\dagger\overleftrightarrow{\p_i}\psi)\!\equiv\!-i/2(\psi^\dagger\p_i\psi -\p_i\psi^\dagger\psi)$}, a spinor rotation $\Psi\to \exp(i/2 V_k \sigma_k)\Psi$ cancels the antisymmetric term leaving behind a term proportional to the divergence of ${\vec V}$. Depending on the character of the vector ${\vec V}$ this term will contribute to the  deformation potential $H_1$ (${\vec V}$ is a vector) or to the pseudoscalar term $b_0$ in ${\tilde H}_{2,0}$ in table \ref{tH} (${\vec V}$ is a pseudovector).

The Hamiltonian $H_{4}$ is a particular case of the previous generic Hamiltonian. 
Using the definition of ${\vec\Omega}$, this term becomes $H=\epsilon_{ijm}\Omega_m \sigma_i k_j$ which can be rotated to $H_{4}={\vec\nabla}\cdot{\vec\Omega}$. Being the divergence of a curl, this term is  zero. Nevertheless, other expressions following the same structure  will produce non--vanishing terms that contribute to $b_0$. 

{\it Strain--independent term and $b_0$ contributions.}
Another interesting example of the previous type of couplings is the strain--independent term linear in $k$ that can accompany the  Weyl Hamiltonian of eq. \eqref{HWSM}: $H_0^a =\epsilon_{ijk}\sigma_ik_j b_k$. When the derivative acts on the spinor fields, this term can be rotated to $H^a_{0}={\vec\nabla}\cdot{\vec b}$ and will induce a pseudoscalar potential similar to $b_0$ at the boundary of the sample where the separation between the nodes  goes to zero. This term breaks inversion symmetry and will induce an energy separation between the nodes. When strain is applied, $b_i\to \hat{u}_{ij} b_j = \hat{A}_i=(u_{ij}+\omega_{ij})b_j$, and the field $b_0^a\sim \partial_i   \hat{A}_i$ will be inhomogeneous through the sample. This contribution to $b_0$ is of higher order in derivatives of the strain than the one described by ${\tilde H}_{2,0}$. 

{\it Pseudo-Zeeman coupling}.
As in the previous (1,0) case, without strain, a Zeeman term can be constructed with the previous Hamiltonian $H^a_{(0)} =\epsilon_{ijk}\sigma_ik_j b_k$ with the derivative acting on the vector ${\vec b}$ separating the Weyl nodes. Although ${\vec b}$ is constant inside the sample, it will always go to zero at the boundary of finite samples giving rise to an effective pseudo-magnetic field confined to the boundary as the one discussed in~\cite{CCetal14}. Consider as an example, a cylinder of WSM of height $L$ and radius $a$ with the simplest configuration $\vec{b}=\hat{z}b_{z}\Theta(a-|r|)$. The corresponding magnetic field will point in the azimuthal direction and be proportional to $B_\theta\sim b_z\delta(a-|r|)$. There will be an associated Zeeman term $H^a_0=\sigma_\theta B_\theta$.

For completeness, we will add the electron--phonon couplings ${\tilde H}_i$ of eq. \eqref{eph} associated to the standard strain tensor $u_{ij}$ defined in eq. \eqref{u}. The lowest order terms are shown in Table \ref{tH}. Some of them  have already been discussed in the literature in the context of WSMs although others are new. The reference where they were first described is shown in the last column of the table.
\begin{widetext}
%
\begin{center}
\begin{table}[h]
\begin{tabular}{|c || c |c | c |c | c |}
\hline
$H_i$ & $(n_q, n_k)$ & Interaction term & Physical interpretation & $K_2$ & Ref.\\
\hline\hline
${\tilde H}_1$ & (0,0) & $\text{Tr}(u)$ & Deformation potential & $+$ &\cite{ACV17}  \\
  ${\tilde H}_{2i}$ & (0,0) & $\sigma^iA_i$, $A_i=u_{ij}b_j$& Pseudogauge field  & $-$ & \cite{CFLV15}\\
${\tilde H}_{2,0}$ & (0,0) & $\sigma^0A_0$ & Pseudoscalar gauge potential & $-$ & \cite{CKLV16}\\
 \hline
${\tilde H}_{3}$ & (0,1) & $k^iA_i$ & Dirac cone tilt & $-$ & New\\
${\tilde H}_{4}$ & (0,1) & $u_{ij} \s_i k_j $& Anisotropic  Fermi velocity & $+$ & \cite{CZ16}\\
 \hline
${\tilde H}_5$ & (1,0) & $\;\;\;\;\epsilon_{ijk}\sigma_ik_jA_k\;\;\;\;$ & Pseudo-Zeeman term & $-$ & New
\\
\hline
\end{tabular}
\caption{Lowest order effective low energy Hamiltonians for the electron-strain interactions around a 3D Weyl node associated to the strain tensor.}
\label{tH}
\end{table}
\end{center}
\end{widetext}

The elastic vector potential described in ${\tilde H}_{2i}$: $A_i=u_{ij}b_j$ was first derived in ref.  \cite{CFLV15} which was followed by a number of works  discussing their physical consequences and performing WSMs straintronics \cite{PCF16,GVetal16,CKLV16,LPF17,ACV17,GMetal17a,GMetal17b,GMetal17c,ZL17,HZS17}. 
This is the simplest term that can written for any ${\cal T}$--broken WSM.  In materials where Weyl points are located at high symmetry points protected by lattice symmetries, other forms of the gauge potentials $H_{2i}$ could arise 
as these discussed in \cite{ZL17}.

The term ${\tilde H}_3$ will tilt the cones of an originally untilted WSM in oposite directions. But strain will also modify a  pre--existing tilt $W_i$  making it inhomogeneous through the sample: 
$W_i\to\tilde W_i(x)=W_i+u_{ij}W_j$. This term opens the possibility of a strain--induced Lifshitz transition, i.e., going from Type-I to Type-II by applying strain. Since the Fermi velocity is also strain--dependent, the condition for the transition  in Eq.~\eqref{1-2}  becomes $\lVert t_i \rVert > 1$, where $t_i = (c A_i + \tilde W_i)(\tilde{v}^{-1})_{ij}$ and $\tilde{v}_{ij}=v_{ij}+u_{ij}$ is the space-dependent Fermi velocity $H_{4}$.

The (1,0) order term ${\tilde H}_5$ can be identified as a pseudo--Zeeman coupling $H_Z={\vec \sigma}\cdot{\vec B^s}$ of the pseudomagnetic field defined in ${\tilde H}_{2i}$: ${\vec B^s}={\vec \nabla}\times{\vec A}$. 

\begin{figure}
\begin{minipage}{0.7\linewidth}
\centering
\includegraphics[width=0.98\columnwidth]{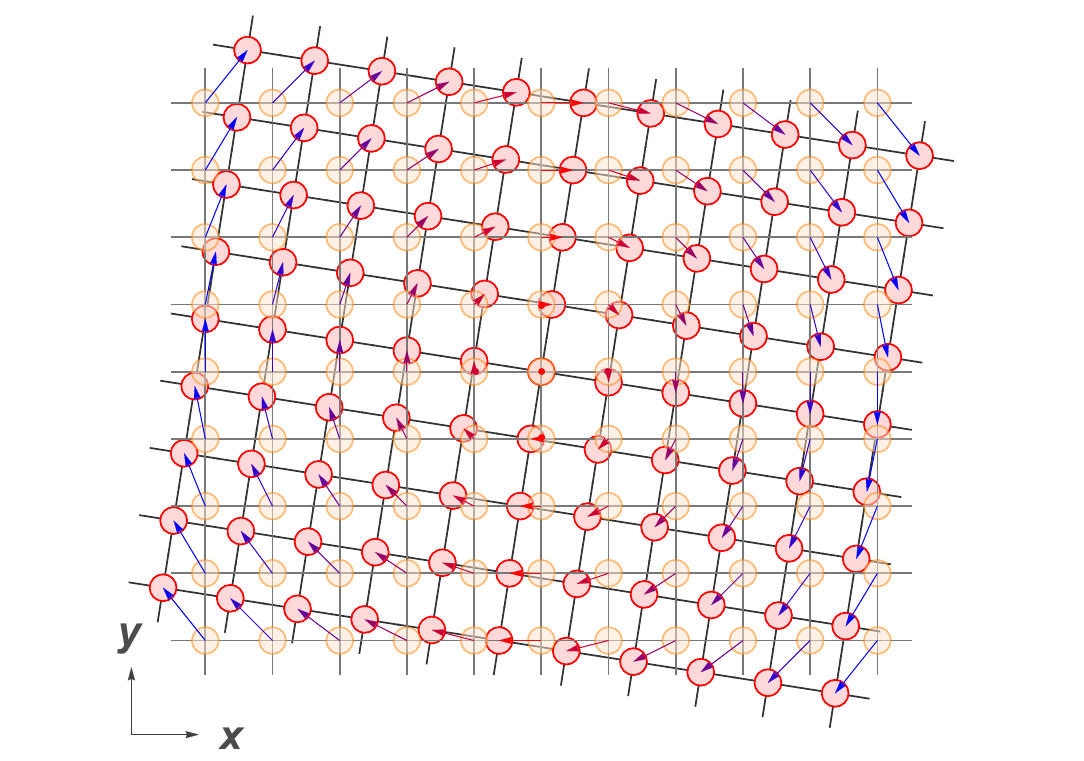}
\end{minipage}
\caption{Schematic representations of the lattice deformation defined in Eq.~\eqref{eq_FD} obtained  by applying torsion to a WSM crystal prepared in a wire geometry as described in the text. } 
\label{Fig_strain}
\end{figure}

{\it Physical example }.
A rotational strain applied to a wire of WSM was discussed in ref.~\cite{PCF16}. The purpose of that work was to discuss pseudo Landau level physics hence the strain constructed was such as to generate a constant pseudomagnetic field of the type ${\tilde H}_{2i}$. We will see the extra terms arising in this situation.  
A tight binding construction provided a material with two Weyl nodes separated in $k_z$ by a vector ${\vec b}=b{\hat u}_z$. Assuming a wire of WSM of length $L$ with an axis along the $z$ direction, applying torsion to the crystal prepared in a wire geometry
the displacement vector in cylindrical coordinates is
\beq
{\vec u}=\theta\frac{z}{L}({\vec r}\times{\hat z}),
\label{eq_FD}
\eeq
where $\theta$ is the twisted angle and ${\vec r}$ denotes the position relative to the origin located on the axis of the wire (a representation of the deformation is shown in the left hand side of Fig.~\ref{Fig_strain}.). It is easy to see that the strain tensor associated to the deformation~\eqref{eq_FD} is traceless, so there will not be a  deformation potential ${\tilde H}_1$. The vector potential ${\tilde H}_{2i}$ and the corresponding uniform pseudomagnetic field are
${\vec A}=\frac{\theta b}{2L}(y, -x, 0),\qquad
{\vec B}^s=-\frac{\theta}{L}b {\hat u}_z$.
The rotation vector ${\vec\Omega}$ of Eq.~\eqref{eq_rot} is given by
${\vec\Omega}=\frac{1}{2} {\vec\nabla}\times{\vec u}=\frac{\theta}{2L}(x, y, -2z)$.
According to our previous discussion, the deformation potential associated to the antisymmetric part of the deformation gradient is
$\Phi(u)={\vec b}\cdot{\vec\Omega}=-\frac{\theta}{L}bz$.
Even though the applied strain is traceless, it is able to produce a deformation potential through the rotation vector. Without time---dependent deformations, this deformation potential will induce an electric field parallel to the pseudomagnetic field ${\vec B}^s$:
${\vec E}^s=-\frac{\partial\Phi}{\partial z}=\frac{\theta}{L}b{\hat u}_z$.
The product ${\vec E^s}\cdot{\vec B^s}$ will not ignite the chiral anomaly because, unlike the pseudomagnetic field $B^s$, the field $E^s$ couples with the same sign to the two Weyl nodes and the contribution from the two nodes cancel. Charge will flow between the two nodes in the presence of a real magnetic field $B_z$ from the anomaly equation \cite{LPQ13}
\beq
\partial_t\rho+{\vec \nabla}\cdot{\vec j}=\frac{e^2}{2\pi^2}({\vec E}\cdot{\vec B}_s)
+({\vec E_s}\cdot{\vec B}),
\eeq
where $E_s$, $B_s$ are strain--induced pseudo--electric and pseudo--magnetic fields.

Due to the contributions of $H_3$, the deformation in Eq.~\eqref{eq_FD} will induce at each node of chirality $\pm$ a tilt of velocity: 
${\vec W}^\pm= \pm [g_1{\vec A}+g_2({\vec b}\times {\vec \Omega}) + g_3 {\vec\Omega}]$,
where $g_i$ are the coupling constants.
It is interesting to note that the total vector potential ${\hat A}_i={\hat u}_{ij}b_j$ would be zero if the  interaction $H_{2i}=\sigma_i A_i$ and the antisymmetric contribution would come with the same coupling constant ($\hat{u}_{i3}$ components are zero). In this case, the only effect of strain would come from the deformation potential and its associated pseudo--electric field discused previously.

The orbital character of the TB models in WSMs makes rotational strain particularly relevant as opposed to the situation in graphene and similar 2D materials having $p_z$ orbitals.
Part of the interest of the emerging axial vector fields associated to strain, lies on the possibility of generating new elastic  or mixed electromagnetic-elastic responses as the ones discussed in the early works \cite{VAetal13,ZJetal13,SHR15}. We have seen an example of this physics in the physical example  where the deformation potential generated from rotational strain in a volume--preserving deformation, generates  an electric field able to ignite the chiral anomaly.

Summarizing, we have shown new electron--phonon couplings associated to rotational strain in three dimensional Dirac matter that enrich the potentials of straintronics, are able to generate new anomaly--related currents, and may induce transitions between different topological semimetals. 

\vspace{0.3cm}
\begin{acknowledgments}
We thank A. Cortijo, F. de Juan, A. Grushin, Y. Ferreiros, B. Amorim, E. V. Castro, J. L. Ma\~nes, and A. Salas for interesting discussions. 
This work  has been supported by Spanish MECD
grant FIS2014-57432-P, the Comunidad de Madrid
MAD2D-CM Program (S2013/MIT-3007), and by the PIC2016FR6. V.A. is supported by an FPI predoctoral contract from MINECO, BES-2015-074072.
\end{acknowledgments}

\bibliography{PRL2}
\end{document}